\documentclass[aps,pra,reprint,amsmath,amssymb,superscriptaddress]{revtex4-1}
\usepackage[utf8]{inputenc}
\usepackage[T1]{fontenc}
\usepackage[english]{babel}

\usepackage{graphicx}
\graphicspath{{pic/}}

\usepackage{hyperref}
\hypersetup{colorlinks=true,linkcolor=blue,citecolor=magenta,breaklinks=true}

\begin{document}

\title{Dielectric particle lofting from dielectric substrate\\ exposed to low energy electron beam} 

\author{P.~V.~Krainov}
\email{pavel.krainov@phystech.edu}
\affiliation{Moscow Institute of Physics and Technology, 
Institutskiy pereulok str. 9, Dolgoprudny, Moscow region 141701, Russia}
\affiliation{Institute for Spectroscopy of the Russian Academy of Sciences, 
Fizicheskaya str. 5, Troitsk, Moscow 108840, Russia}

\author{V.~V.~Ivanov}
\affiliation{Institute for Spectroscopy of the Russian Academy of Sciences, 
Fizicheskaya str. 5, Troitsk, Moscow 108840, Russia}

\author{D.~I.~Astakhov}
\affiliation{ISTEQ B.V., High Tech Campus 9, 5656 AE Eindhoven, The Netherlands}
\affiliation{Institute for Spectroscopy of the Russian Academy of Sciences, 
Fizicheskaya str. 5, Troitsk, Moscow 108840, Russia}

\author{M.~A.~van~de~Kerkhof}
\affiliation{ASML Netherlands B.V., De Run 6501, 5504DR Veldhoven, The Netherlands}

\author{V.~V.~Kvon}
\affiliation{ASML Netherlands B.V., De Run 6501, 5504DR Veldhoven, The Netherlands}

\author{V.~V.~Medvedev}
\affiliation{Institute for Spectroscopy of the Russian Academy of Sciences, 
Fizicheskaya str. 5, Troitsk, Moscow 108840, Russia}
\affiliation{Moscow Institute of Physics and Technology, 
Institutskiy pereulok str. 9, Dolgoprudny, Moscow region 141701, Russia}

\author{A.~M.~Yakunin}
\affiliation{ASML Netherlands B.V., De Run 6501, 5504DR Veldhoven, The Netherlands}

\date{\today}

\begin{abstract}
The particle-in-cell simulation is applied to study a nanometer-sized dielectric particle lofting 
from a dielectric substrate exposed to a low energy electron beam.
The article discusses the electron accumulation between such a substrate 
and a particle lying on it,
that can cause a particle lofting.
The results are of interest for dust mitigation in the semiconductor industry, 
the lunar exploration and the explanation of the dust levitation. 
\end{abstract}


\maketitle 

\section{Introduction\label{sec:intro}}
Extreme ultraviolet lithography (EUVL) is a technology for integrated circuits (IC) manufacturing \cite{vandekerkhof_2019}.
This technology uses EUV light of 13.5 nm wavelength to transfer a pattern from 
a photomask (also called reticle) to a light-sensitive photoresist on a wafer \cite{wagner_2010}. 
In view of the IC feature sizes of < 20 nm, any particles on the surface of a reticle of > 20 nm 
will cause defective patterns to be printed \cite{scaccabarozzi_2009}. 
Therefore, control of release and transport of these nanoparticles is vitally important for EUVL \cite{lercel_2019}.

The process of EUVL \cite{beckers_2019} occurs in a low pressure hydrogen atmosphere
to prevent oxidation of mirrors and carbon growth.
The absorption of EUV radiation results in EUV induced hydrogen plasma formation.
It consists of two parts:
fast photoelectrons ($E\ \sim$ 70 eV) 
and a bulk plasma ($n_e\ \sim\ 10^8\ \text{cm}^{-3}$, $T_e\ \sim\ 0.5$ eV).
Both the fast electrons and the plasma charge surfaces they can reach.
It has been reported in several experiments \cite{sheridan_1992,flanagan_2006,wang_2016} 
that a plasma and an electron beam with similar parameters
can lift off dust particles from surfaces.

In 1992, Sheridan et al. \cite{sheridan_1992} observed the shedding of the dielectric dust 
from an aluminum sphere covered by an oxide layer and simultaneously exposed to a plasma and an electron beam.
According to the reported hypothesis, that was expanded later \cite{sheridan_2011},
a particle is charged by the plasma and lifted by the electric field of the plasma sheath.
In 2006, Flanagan and Goree \cite{flanagan_2006} repeated Sheridan's experiment
for a glass sphere covered with the regolith and got the same dust shedding.    
Wang et al. \cite{wang_2016} investigated the lofting from a heap of regolith particles 
under the influence of a plasma, an electron beam, their combination and UV radiation.
In accordance with the developed "patched charge model", electrons penetrate 
into cavities between particles, 
charge their hidden surfaces with the help of secondary electron emission
and causes the release.
All of the aforementioned authors noted the key importance of an electron beam addition in a plasma 
for the lofting phenomenon.

The core objective of this article is to show a new mechanism of 
a single particle lofting from a surface.
Here we deal with a single dielectric particle lying on a dielectric substrate 
exposed to a low energy electron flux, 
without a plasma or high energy EUV photons. 
By kinetic numerical simulation, we demonstrate the accumulation of electrons 
between a particle and a substrate,
that leads to the occurrence of a high repulsive force during charging transient.
Besides, the authors give the consistent description of particle charging,
including a dielectric mirror force,
that was not considered previously in literature, to the best of our knowledge.

\section{Model and geometry\label{sec:model}}
The process of a particle and a substrate charging was simulated by means of the
2-dimensional (r-z axial symmetry) model \cite{astakhov_2016}
based on a particle-in-cell method \cite{charleskbirdsall_1991}
and developed for the simulation of EUV induced plasma.

The lower part of the simulation domain was filled by a dielectric substrate.
Its thickness was equal to 2 $\mu m$ in order to avoid an interaction between the particle 
and its image in the grounded metal substrate below the dielectric substrate.
A spherical dielectric particle (100 nm) was placed $z\ =\ 1$ nm above the substrate
to account for the surface roughness.
But the shape of both the particle and the substrate were smooth before projection on a grid,
otherwise too dense grid was needed to resolve all details.
It was assumed that the particle and the substrate 
were made of silicon dioxide ($\varepsilon\approx 3.9$). 

The substrate was exposed to a spatially and temporally uniform flux 
of electrons (70 eV) $flux\ =\ 10^{12}\ cm^{-2}s^{-1}$.
A vertical incidence of electrons was implemented in order to exclude the contribution of
primary electrons to the accumulation of electrons under the particle and 
focus on the accumulation due to secondary electrons from the substrate.
No external electric field was applied because of expected negligible influence.

The simulation domain boundaries were placed > 5 $\mu m$ away from the particle.
The potential of a charged particle equals to about zero at such a distance.
Therefore, all boundaries could be considered as infinitely far surfaces
and set as grounded electrodes. 
Non-uniform rectangular grid was exploited to solve the Poisson equation.
A cell size in the place of particle location was 1 nm. 

The following dielectric properties were taken into the consideration: 
the conservation and the accumulation of charges;
the polarization according to the preliminary chosen dielectric permittivity;
the electron backscattering and 
the secondary electron emission induced by electrons (SEE).
Two last processes were merged into one process of SEE
due to the lack of experimental data
that did not allow to separate those processes entirely.
The characteristics of SEE included secondary electron emission 
yield (inelastic backscattering \cite{dunaevsky_2003}, 
"true" secondary emission \cite{lin_2005,ooka_1973}),
the spectrum of secondary electrons from silicon dioxide \cite{schreiber_2002}
and angular distribution \cite{bundaleski_2015}.

\section{Attractive and repulsive forces\label{sec:repulsive_force}}
The bombardment of a flat dielectric substrate by 70 eV electrons 
leads to a positive charging of the substrate,
because released secondary electrons leave it freely and secondary emission yield $SEY(70\ eV)\ >\ 1$.
The substrate acquires a low positive potential $\phi\ \sim\ 1\ V$,
because most secondary electrons are cold 
(the spectrum has maximum point at $E_{sec. max}\ \approx\ 3\ eV$ \cite{schreiber_2002}).

\begin{figure}[t]
  \includegraphics[width=\linewidth]{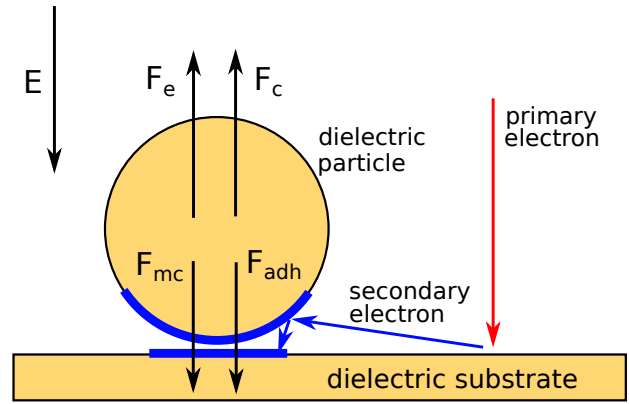}
  \caption{The mechanism of electron accumulation. 
  A primary electron hits the substrate.
  It releases a secondary electron, that hits the particle. 
  It can release another secondary electron (short blue arrow), 
  which hits the substrate, and so on.
  Forces: 
  $F_e$ - interaction with the electric field $E$, 
  $F_c$ - Coulomb repulsion by electrons under the particle,
  $F_{adh}$ - adhesive force,
  $F_{mc}$ - mirror force.\label{fig:accumulation}}
  
\end{figure}

Some of secondary electrons released next to a particle
hit it, then experience either absorbation, or backscattering, or release new secondary electrons.
New secondary electrons in their turn hit the substrate under the particle, 
undergoing the same process again and again.
The described process leads to the accumulation of electrons locally on the particle surface 
and on the surface of the substrate under the particle as is shown on fig.~\ref{fig:accumulation}.

The accumulation of electrons under the particle results in 
an extra repulsive force $F_c$ (fig.~\ref{fig:accumulation}),
because of Coulomb repulsion between electrons. 
Besides, there are four more forces acting on the particle (fig.~\ref{fig:accumulation}).

In general, the particle might be influenced by an external electric field.
It might be a field of plasma sheath or a specially applied external field. 
Force $F_e$ - action of external field $E$ on the total particle charge $Q_p$.
\begin{equation}
  F_e = Q_p\cdot E \text{.}
\end{equation}
This force was supposed in previous articles \cite{sheridan_1992,flanagan_2006,sheridan_2011,heijmans_2017} 
to be responsible for the particle lofting.

Van der Waals force is usually considered as adhesive force $F_{adh}$ in vacuum.
For example, an attraction between a smooth spherical particle and a smooth plane equals to
\begin{equation}
  F_{adh} = F_{VdW} = \frac{A_{ham}}{6} \cdot \frac{R}{z^2} \text{,}
\end{equation}
where R - the particle radius, z - the separation distance between the particle and the surface, 
$A_{ham}$ - the Hamaker constant ($0.66\cdot 10^{-21}$ J for SiO$_2$ \cite{israelachvili_2011}).

A dielectric is polarized in an electric field
and thus forms an electrical image for any charge above it.
For instance, a point-like charge $q$ placed in vacuum at the distance $L$ 
from a flat dielectric with permittivity $\varepsilon$
experiences the attraction to the surface with a force 
\begin{equation}
  F_{mc\ point-like} = \frac{\varepsilon - 1}{\varepsilon + 1}\frac{q^2}{(2L)^2} \text{.}
\end{equation}
Thus, the particle is attracted to the substrate with the mirror force $F_{mc}$.
It includes 
the interaction of charges on the particle with their images in the dielectric substrate and
vice versa.

Gravitational force (it is not shown in fig.~\ref{fig:accumulation}) is negligible 
as it is several orders lower than e.g. Van der Waals force 
considered in this article.

Total force acting on the particle
\begin{equation}
    F_t = F_e + F_C - F_{mc} - F_{adh} \text{.}
\end{equation}

\begin{figure*}[ht]
  \begin{minipage}[h]{0.49\linewidth}
  \center{\includegraphics[width=\linewidth]{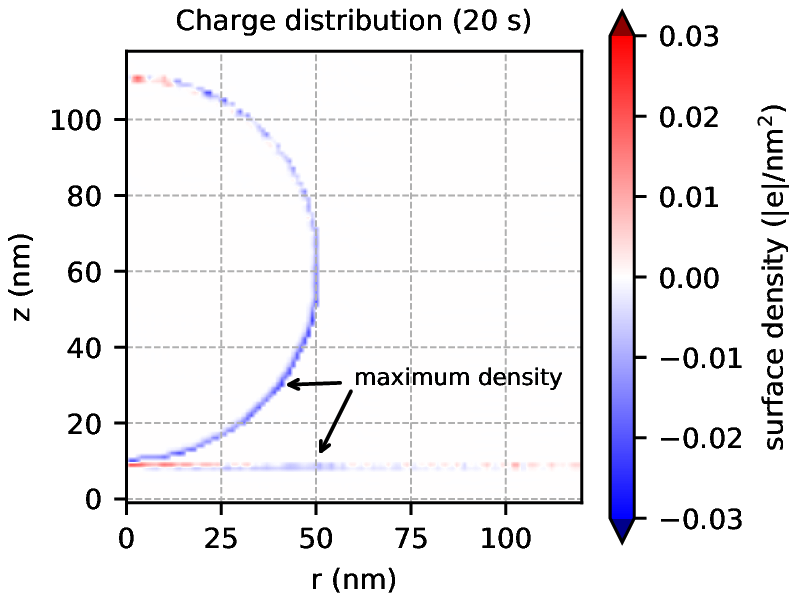}}
  \end{minipage}
  \hfill
  \begin{minipage}[h]{0.49\linewidth}
  \center{\includegraphics[width=\linewidth]{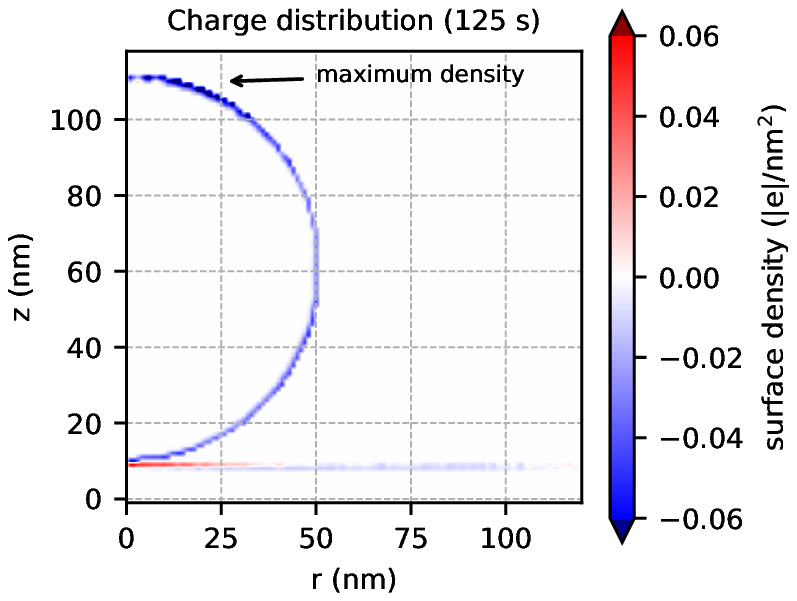}}
  \end{minipage}
  \caption{Charge distribution over the surfaces of a dielectric particle 
  and a dielectric substrate exposed to the 70 eV electron flux.
  Negative charge density under the particle shows electron accumulation.
  Positive charges under the particle are polarization charges 
  induced by negatively charged particle.
  In the left picture charge density reaches its maximum value under the particle, 
  leading to repulsive force (period A in fig.\ref{fig:force_vs_time}).
  In the right picture charge density reaches its maximum value 
  on the upper part of the particle, leading to attraction due to mirror force 
  (periods B and C in fig.\ref{fig:force_vs_time}). \label{fig:dielectric_charge_density}}
  
\end{figure*}

\section{Results\label{sec:results}}
The charge distributions on the particle and the substrate obtained in the simulation
are shown in fig.~\ref{fig:dielectric_charge_density}.
In the very first stage of particle charging (fig.~\ref{fig:dielectric_charge_density}) 
the area with the highest charge density forms on the lower part of the particle.
Also, there is an area with negative charges on the substrate.
Its charge density decreases with an increase of the distance from the particle.
Both these results demonstrate the accumulation of electrons under the particle
and provide conditions for occurrence of a strong repulsive force.
A positive charge density is formed in the place of a contact between the particle and the substrate.
It can be explained by polarization charges induced by the negatively charged particle.

\begin{figure}[t]
  \includegraphics[width=\linewidth]{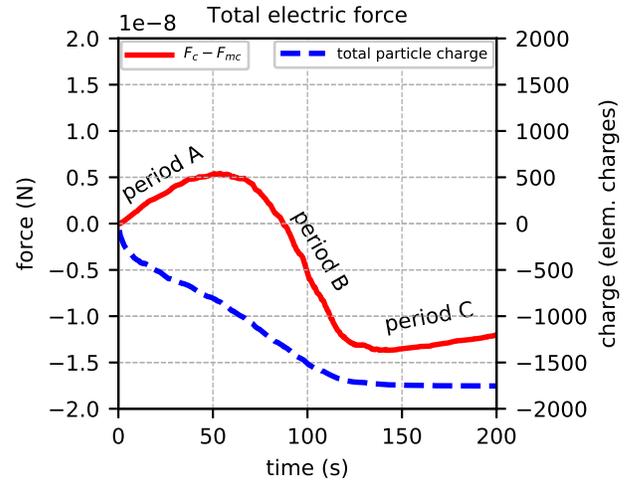}
  \caption{Total electric force ($F_{c} - F_{mc}$) 
  acting on the particle
  and total particle charge.
  A: the growth of repulsion force due to the electron accumulation under the particle;
  B: the rise of mirror force due to the fast charging of the upper part of the particle;
  C: electron accumulation under the particle continues slowly.\label{fig:force_vs_time}}
  
\end{figure}

The total electric force acting on the particle is presented in fig.~\ref{fig:force_vs_time}.
It includes two parts: mirror force $F_{mc}$ and Coulomb repulsion $F_{c}$.
The third force $F_e$ can be neglected because no external electric field was applied and 
the potential difference formed during the simulation
between the substrate surface and the upper boundary does not give a significant force.
The volume charge did not form in the simulation domain 
because of the weak electron flux and the small size of the domain.

The total electric force has a transient nature, 
because different parts of the particle are charged with various rates 
by different kinds of electrons (primary or secondary).
The following sequence was established.
During the period A (fig.~\ref{fig:force_vs_time}), the lower part of the particle 
is negatively charged by low energy secondary electrons 
(for most of them \cite{schreiber_2002} $E_{sec.}\ \sim\ 3\ eV$) released from the substrate. 
Secondary emission yield $\overline{SEY(E_{sec.})} < 1$ for such electrons,
that makes a negative charging possible.
The lower part can be charged 
up to about -60~V because of a long "tail" of 
the secondary electrons spectrum \cite{schreiber_2002}.
During the period A, the upper part of the particle 
is charged a bit positively (fig.~\ref{fig:dielectric_charge_density}a)
by primary electrons, because $SEY(E_{pr.}) > 1$.
The total electric force acting on the particle rises during this period 
due to the electron accumulation under the particle 
(fig.~\ref{fig:force_vs_time}: 0-60 seconds).

During the period B (fig.~\ref{fig:force_vs_time}), the accumulated electrons 
induce a negative potential in the upper part of the particle.
Therefore primary electrons hit the upper part with less energy ($E_{pr.}^{'}$).
At some point in time, the upper part of the particle starts charging,
because $SEY(E_{pr.}^{'})\ <\ 1$.
The mirror force acting on the particle increases and the total electric force changes its sign
after about 80 seconds.
In other words, the described accumulation of electrons under the particle acts 
as a trigger for the upper part of the particle.
It charges positively first, and after the trigger negatively.
Such a mechanism is responsible for the flip and transient nature of the total electric force.

In the period C, the lower part slowly continues charging, 
whereas the upper part is saturated with charges.

\section{Discussion\label{sec:discussion}}

This paragraph is aimed to discuss the reasoning of the chosen 2-D r-z model. 
Unlike a hypothetical 3D model, 
2D r-z model cannot take interaction between 
single point-like charges into account in the right way,
because a single point-like charge is represented in the model as a ring around the axis of symmetry.
Nevertheless, the dielectric particle got more than 1000 electrons during its charging
as fig.~\ref{fig:force_vs_time} demonstrates.
These electrons should be uniformly distributed around the symmetry axis of a particle,
because of axial symmetry of the problem.
Therefore, 2D model should give the same value of the electric force as a hypothetical 3D model.

\begin{figure}[t]
  \includegraphics[width=\linewidth]{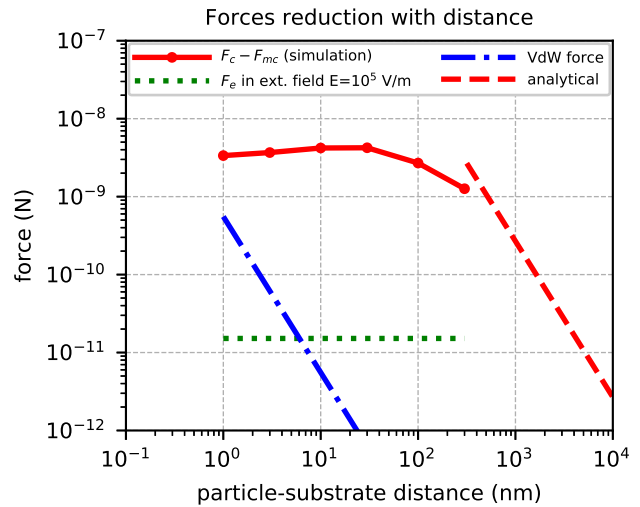}
  \caption{Reduction of forces acting on the particle with particle-substrate separation distance.
  Charge distribution corresponds the maximum value of total electric force 
  (60 sec. in fig.~\ref{fig:force_vs_time}).
  Van der Waals force includes only non-contact term, i.e. interaction between main bodies.
  Analytical approach considers the particle and the electron spot on the substrate as points
  and employs Coulomb law to calculate the force. \label{fig:force_vs_distance}
           }
  
\end{figure}

The main idea of fig.~\ref{fig:force_vs_distance} is to show
that accumulation of sufficient amount of electrons under the particle 
leads to a subsequent particle lofting.
For this purpose, we fixed the charge distribution on the particle and on the substrate
and calculated the total electric force 
for several particle-substrate separation distances with the help of our model 
(red dots in fig.~\ref{fig:force_vs_distance}). 
A corresponding distribution of polarization charges was calculated for
every new separation distance.
We used the distribution of charges, 
that corresponded to the maximum value of the total electric force,
because of Van der Waals force uncertainty. 
It was described in e.g. \cite{lamarche_2017},
that contact term caused by surfaces roughness can exceed 
non-contact term of Van der Waals force (i.~e. interaction between main bodies)
one order or even more.
Fortunately, such an uncertainty can be compensated by a large difference 
between obtained values of the total electric force 
and estimated non-contact Van der Waals force (fig.~\ref{fig:force_vs_distance}).

It was implicitly assumed that the particle charge distribution does not change during its lofting.
It can be supported by two evaluations.
The time between two subsequent electron-particle collision is
\begin{equation}  
    \tau_e = (flux \cdot \pi R^2)^{-1} \approx 6\ \text{ms.} 
\end{equation}
To estimate a typical time of lofting assume the particle diameter as characteristic length
and value of Van der Waals force as characteristic force, then:
\begin{equation}
  \tau_r = \sqrt{\frac{2R}{F_{VdW}/m}} \approx 10\ \text{ns} \ << \tau_e \approx 6\ \text{ms.}
\end{equation}

Authors suppose a similar electron accumulation to be possible 
if a particle is simultaneously exposed to a plasma and a flux of fast electrons.
For the sake of comparison, the green line in fig.~\ref{fig:force_vs_distance} shows a force
that acts on the particle in external field $E\ =\ 10^5$ V/m. 
The given value is the estimation of a maximum electric field in a plasma sheath of EUV induced plasma.
Therefore, particle jump on about 10 nm above the surface results in a significant reduction of adhesion.
Van der Waals force becomes less than force in electric field $E$ and 
the following lofting can be driven by this electric field.
It can carry the particle through the sheath to the bulk plasma.

The described lofting could occur for a metal particle on a metal substrate.
Their surfaces can be partly covered by a naturally formed surface oxide layer with a width of several nm,
that can make the described electron accumulation possible.
But in such conditions mirror force is not weakened by $(\varepsilon - 1)/(\varepsilon + 1)$ coefficient
and has more significant effect on a particle.

\section{Conclusion\label{sec:conclusion}}
The article demonstrated a new mechanism of a dielectric particle lofting
from a dielectric substrate exposed to a low energy electron beam.
By the particle-in-cell simulation, the charging transient of the particle was investigated.
We revealed the effect of considerable electron accumulation
between such a particle and a substrate,
leading to the occurrence of a strong repulsive force acting on the particle.
The accumulation is followed by the rise of the total particle charge
that causes the particle attraction to the substrate due to an increased mirror force.
The obtained repulsive force can exceed an attractive adhesive force.
This repulsive force reduces slower with the particle-substrate separation distance than the adhesive force.
It results in the particle lofting as soon as the sufficient amount of electrons has been accumulated.

\bibliography{particle-lofting-article-en.bib}

\end{document}